# ARTICLE

# Valence-shell photoelectron circular dichroism of ruthenium(III)-tris-(acetylacetonato) gas-phase enantiomers

Benoît Darquié,[*a] Nidal Saleh,[b,c] Sean K. Tokunaga,[a] Monika Srebro-Hooper,[*d] Aurora Ponzi,[e,f] Jochen Autschbach,[g] Piero Decleva,[*e] Gustavo A. Garcia,[h] Jeanne Crassous,[*b] Laurent Nahon[*h]

Chiral transition-metal complexes are of interest in many fields ranging from asymmetric catalysis and molecular materials science to optoelectronic applications or fundamental physics including parity violation effects. We present here a combined theoretical and experimental investigation of gas-phase valence-shell photoelectron circular dichroism (PECD) on the challenging open-shell ruthenium(III)-tris-(acetylacetonato) complex, Ru(acac)$_3$. Enantiomerically pure $\Delta$ or $\Lambda$-Ru(acac)$_3$, characterized by electronic circular dichroism (ECD), were vaporized and adiabatically expanded to produce a supersonic beam and photoionized by circularly-polarized VUV light from the DESIRS beamline at Synchrotron SOLEIL. Photoelectron spectroscopy (PES) and PECD experiments were conducted using a double imaging electron/ion coincidence spectrometer and compared to density functional theory (DFT) and time-dependent DFT (TDDFT) calculations. The open-shell character of Ru(acac)$_3$, which is not taken into account in our DFT approach, is expected to give rise to a wide multiplet structure, which is not resolved in our PES signals but whose presence might be inferred from the additional striking features observed in the PECD curves. Nevertheless, the DFT-based assignment of the electronic bands leads to the characterisation of the ionized orbitals. In line with other recent works, the results confirm that PECD persists independently on the localization and/or on the achiral or chiral nature of the initial orbital, but is rather a probe of the molecular potential as a whole. Overall, the measured PECD signals on Ru(acac)$_3$, a system exhibiting $D_3$ propeller-type chirality, are of similar magnitude compared to those on asymmetric-carbon based chiral organic molecules which constitute the vast majority of species investigated so far, thus suggesting that PECD is a universal mechanism, inherent to any type of chirality.

## Introduction

Chirality of metal complexes has been continuously attracting attention in fundamental and applied areas of coordination chemistry. For instance, metal-tris(β-diketonato) complexes are chiral octahedral metal $D_3$-symmetric molecules with intrinsic chirality due to their typical propeller-like structure and display the classical $\Delta/\Lambda$ stereochemistry.[1,2] M(acac)$_3$ (where M is a metal and acac stands for acetylacetonato) is the archetype of such complexes and has been used as a chiral model for chiroptical spectroscopies[3-5] and in applications such as in catalysis,[6] or in chiral liquid crystals.[7-10] Chiral resolution and racemization of metal-tris(β-diketonato) complexes have been studied extensively.[11-13]

Another appealing aspect of these complexes is their relatively high volatility which has proven very useful to prepare metallic surfaces for catalytic and optoelectronic applications through sublimation and chemical vapour deposition.[14-16] This particular feature makes them also ideally suited for investigating chirality-related fundamental properties *via* gas-phase spectroscopy. One such study is the measurement of the expected tiny energy difference between enantiomers induced by the parity violation (PV) inherent to the weak force, which can be probed using high-precision mid-infrared molecular beam spectroscopy.[17-19] Measuring PV in molecules is interesting for a wide range of subjects, from probing the weak force and the limits of the standard model of particle physics to stimulating research on the origins of biomolecular asymmetry.

[a.] *Laboratoire de Physique des Lasers, Université Sorbonne Paris Nord, CNRS, Villetaneuse, France. Email: benoit.darquie@univ-paris13.fr*
[b.] *Univ Rennes CNRS, ISCR—UMR 6226 ScanMat – UMS 2001, 35000 Rennes, France. Email: jeanne.crassous@univ-rennes1.fr.*
[c.] *Current address: Department of Organic Chemistry, University of Geneva, Quai Ernest Ansermet 30, 1211 Geneva 4, Switzerland.*
[d.] *Faculty of Chemistry, Jagiellonian University, Gronostajowa 2, 30-387 Krakow, Poland. Email: srebro@chemia.uj.edu.pl.*
[e.] *CNR IOM and Dipartimento di Scienze Chimiche e Farmaceutiche, Universita' di Trieste, I-34127 Trieste, Italy. Email: decleva@units.it.*
[f.] *Current address: Department of Physical Chemistry, Ruđer Bošković Institute, 10000 Zagreb, Croatia.*
[g.] *Department of Chemistry, University at Buffalo, State University of New York, Buffalo, NY 14260, USA.*
[h.] *Synchrotron SOLEIL, L 'Orme des Merisiers, St. Aubin, BP 48, 91192 Gif sur Yvette, France. Email: laurent.nahon@synchrotron-soleil.fr.*

Electronic Supplementary Information (ESI) available: [details of any supplementary information available should be included here]. See DOI: 10.1039/x0xx00000x

In this context, M(acac)$_3$ compounds happen to be particularly promising, since chiral organo-metallic complexes bearing a heavy metal have been predicted to show notably large parity-violating energy differences.[17,20-24]

**Scheme 1.** Chemical structure of $\varDelta$- and $\varLambda$-Ru(acac)$_3$

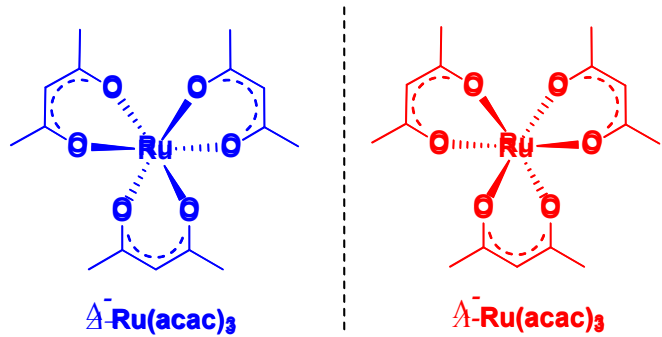

The ability to bring such large chiral compounds into the gas phase makes it also possible to study their photoelectron circular dichroism (PECD), an orbital-specific chiroptical effect involving molecular photoionisation, and thus the electronic continuum.[25] As such, PECD is a precious probe of electronic structure, a crucial aspect for the study of organo-metallic compounds. More generally, within the last two decades, PECD has also become a well-established probe of molecular chirality, being observed as a very intense forward-backward asymmetry in an angle-resolved measurement of the photoelectron yield resulting from the circularly polarized light (CPL)-induced photoionization of pure gas-phase enantiomers. More precisely, for randomly oriented molecules the normalized photoelectron angular distribution function is written as $I_p(\theta) = 1 + b_1^{\{p\}} P_1(\cos\theta) + b_2^{\{p\}} P_2(\cos\theta)$ with $P_j$ being the Legendre polynomial of order j, $\theta$ - the direction of the emitted electron and $p$ - the polarization of the ionizing radiation ($p$ = 0, +1 and -1 for linear, left circular and right circular polarizations, respectively). For CPL, $\theta$ is measured from the photon propagation axis. The $b_1$ parameter is nonzero only for chiral systems photoionized with CPL, and thus encapsulates the chiral contribution, and the first Legendre polynomial term describes a forward-backward asymmetry in the electron emission direction. The $b_1^{\{\pm 1\}}$ coefficient is antisymmetric under swapping of either light helicity or enantiomer, while $b_2$ does not vary under such exchanges. The PECD is defined as $2b_1^{\{+1\}} P_1(\cos\theta)$, the difference between the normalized angular distributions obtained with left and right circular polarization radiations, corresponding to a maximum asymmetry of $2b_1^{\{+1\}}$ in the forward-backward direction.

The PECD-induced asymmetry originating in the scattering of the outgoing photoelectron off an intrinsically asymmetric potential appears to be a very sensitive probe of molecular structures including chemical substitutions, isomers, conformers, clustering and vibrational motion (for recent reviews, see refs. [26-28]). While PECD has been investigated for a variety of organic chiral species such as terpenes,[29-34] oxirane derivatives[35-42] and amino-acids,[43-46] all exhibiting central chirality resulting from asymmetric carbons, to our knowledge, only one organo-metallic complex, the closed-shell system cobalt(III)-tris-(acetylacetonato), Co(acac)$_3$, has been studied so far with this technique. This complex, which exhibits a propeller-like chirality, was examined in enantiopure form by Catone *et al.*[47] who observed a resonant-PECD signal upon relaxation of the 3p → 3d core excitation into the highest occupied molecular orbital (HOMO) channel ionization. In recent years, attempts have also been made to supplement experimental studies with quantum-chemical simulations of PECD, and although a good measure of success has been met in organic molecules, extension to organo-metallic systems remains uncertain. In fact, after a reasonable description of PECD relative to the HOMO ionization in Co(acac)$_3$, further results on deeper ionizations were found to be quite disappointing.[47] It is therefore important to extend those studies on transition-metal-based complexes, to accumulate results and build-up experience on the adequacy and shortcomings of the theoretical tools in order to improve them.

Herein, we report an experimental and theoretical study focused on PECD of enantiopure Ru(acac)$_3$ (see Scheme 1) that interestingly exhibits quite delocalized orbitals. With its chirality coming from the helical arrangement of the three non-chiral ligands around the metal, it offers an interesting new benchmark for PECD to address the still open question of the role of the outgoing electron initial orbital localization with respect to the chiral centre: will the scattering phase of the outgoing electron be mainly sensitive to the nearby presence of a chiral centre (on the metal)? To what extent could PECD be a more global probe of the whole (chiral) molecular potential by exhibiting a strong PECD asymmetry also for pure ligand-based orbitals? Will PECD be as sensitive to propeller-type chirality as compared to asymmetric carbon-based central chirality? These are important questions regarding the general applicability of PECD in an analytical context as being a probe of the whole molecular, and possibly supra-molecular, potential.

Conversely, to study possible localization effects, one needs a precise description of the complex electronic structure of Ru(acac)$_3$, which in turn can be facilitated by exploiting PECD as a sensitive probe of the electronic initial orbitals within a close experience/modelling interplay. The open-shell electronic structure of Ru(acac)$_3$ is however quite challenging to describe. In an idealized $D_3$ geometry, the ground state is a degenerate e$^3$ HOMO configuration, so that valence-shell ionization gives rise to several electronic states, and the one-to-one correspondence between orbitals and ionic states, or photoelectron bands, is lost.[48] This will basically apply also to the small symmetry breaking in the actual geometry.

The paper is organized as follows. The next Section (*Materials and methods*) is devoted to a full description of the various experimental and theoretical methods we used. In Section *Results and discussions*, we present experimental electronic circular dichroism (ECD) results followed by Photoelectron spectroscopy (PES) and PECD measurements obtained by double imaging electron/ion coincidence spectroscopy (i$^2$PEPICO), which are then discussed in light of PES- and PECD-dedicated density functional theory (DFT) and time-dependent DFT (TDDFT) calculations.

## Materials and methods

### a. Ru(acac)$_3$ synthesis

Large quantities (grams) of pure $\Delta$ and $\Lambda$ enantiomers of Ru(acac)$_3$ were synthesized for this work following a known procedure,[11] *i.e.* by mixing 2.4 g of *rac*-Ru(acac)$_3$ (6 mmol, 1 equivalent) with 5.6 g of L-(-)-dibenzoyl-tartaric acid monohydrate (15 mmol, 2.5 equivalents) in 195 mL of a 1:2 benzene/cyclohexane solvent mixture. After stirring for two days, the precipitate (containing the $\Delta$ isomer) was separated from the mother liquors (containing the $\Lambda$ isomer). The precipitate was redissolved in dichloromethane and treated with aqueous sodium bicarbonate (NaHCO$_3$), to yield $\Delta$-Ru(acac)$_3$ (1.08 g, 45% yield). The mother liquors were dried, redissolved in dichloromethane and treated with aqueous sodium bicarbonate, yielding $\Lambda$-Ru(acac)$_3$ (1.15 g, 46% yield). The stereochemistry and enantiomeric purity were checked by comparing the ECD spectra (*vide infra*) with published ones. Note that we cannot use optical rotation (OR) as a characteristic chiral property in this case, due to the absorbance of the compound in the wavelengths range that is typically used in OR measurements (such as for example sodium D-line).

### b. Photoelectron spectroscopy and photoelectron circular dichroism measurements

The PES and PECD measurements have been conducted at the undulator-based, variable polarization VUV beamline DESIRS,[49] at Synchrotron SOLEIL. Owing to a gas filter, the beamline delivers harmonic-free, quasi-perfect CPL with more than 97% circular polarization rate as measured *via* a dedicated polarimeter.[50] This beamline is connected to a versatile continuous molecular beam chamber, called SAPHIRS, equipped with a two-stage differential pumping system.[51] A molecular beam of enantiomerically pure Ru(acac)$_3$ was formed by expanding through a nozzle, the corresponding vapour produced in a stainless steel oven (at temperatures in the 150-200°C range) diluted in 0.5 bars of He. Various nozzle diameters ranging from 50 to 200 µm have been used throughout three different experimental runs, resulting in consumptions of less than 20 mg/hour. The supersonic expansion was then skimmed by two consecutive skimmers (1 mm and 2 mm diameter) before crossing, at a right angle, at the center of the i$^2$PEPICO spectrometer DELICIOUS3,[52] the ionizing VUV CPL from DESIRS.

The resulting electrons and ions were accelerated by a continuous electric field, respectively towards the Velocity Map Imaging spectrometer and a modified Wiley-McLaren 3D momentum imaging mass spectrometer. The extraction field was set to accept, at a given photon energy, all electrons and ions independently of their momenta. The electron-ion coincidence scheme was used here to deliver mass-selected photoelectron images free of any background or spurious electron signal due to impurities or decomposition products. In addition, photoelectron images were further filtered by ion impact position, where only ions falling within a chosen region of interest (ROI) were taken into account. The ROI was chosen to select only ions possessing a net velocity along the molecular beam, which greatly enhances the sensitivity and signal-to-noise ratio (SNR) of the experiment.

From the 3D momentum of the parent Ru(acac)$_3$ ions the translational energy distribution can be obtained (see Fig. S1 of the ESI). From this distribution we estimate a translational temperature of 80 K. We stress, however, that the adiabatic expansion is not in thermal equilibrium so that the internal temperature is harder to estimate. For a given enantiomer, full i$^2$PEPICO data were recorded by alternating right-handed (RCP) and left-handed circular polarization (LCP) light, swapped every 10-15 minutes. The ion mass- and position-filtered photoelectron images provided, after Abel inversion *via* the pBasex algorithm,[53] the $b_1$ parameter from the (LCP-RCP) difference image, while the (LCP+RCP) sum image gave the corresponding photoelectron spectrum according to a well-established procedure.[31] We have carried out both PES and PECD measurements at photon energies ranging from 7.1 eV to 18 eV.

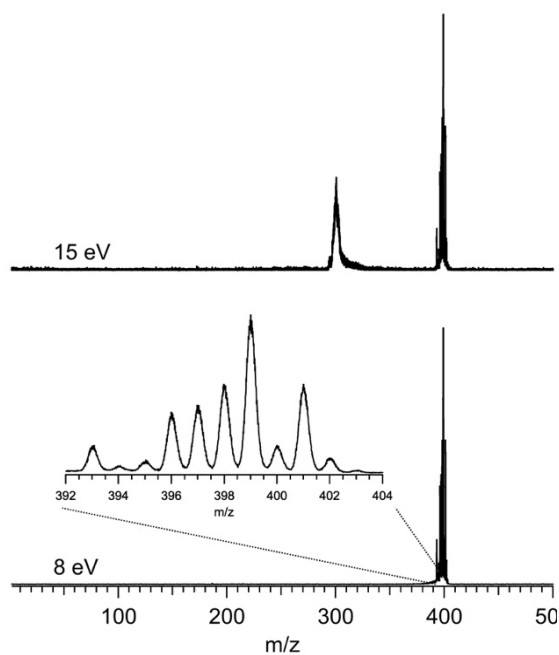

**Fig. 1.** Mass spectra of Ru(acac)$_3$ recorded at 8 eV and 15 eV photon energies. Only a region of the ion detector corresponding to ions coming from the molecular beam is processed for making these graphs. A zoom on the isotopic distribution due to the ruthenium atom is offered in the inset.

Fig. 1 shows two typical mass spectra recorded at low (8 eV) and high (15 eV) photon energies. To avoid noise and spurious signal from background molecules such as water or oxygen, the data from the ion detector were ROI-filtered and Fig. 1 displays only the contributions of ions coming from the molecular beam which, given their initial velocity in the beam, fall on a specific region of the detector (see Fig. S2 of the ESI for details of the ROI-filtering effects). At low-photon energies the only visible species is the parent Ru(acac)$_3$ cation, exhibiting a broad isotope distribution, centred at m/z 399. As discussed in Subsection b of Section *Results and discussions*, dissociative ionisation of parent Ru(acac)$_3$ starts at around 12 eV and corresponds to the loss of one ligand to produce m/z 300. No other fragments are seen in the photon energy range studied.

The mass spectra also show that no decomposition or spurious synthesis compounds are present in the beam. However, in order to improve the SNR especially at high-photon energy, the PES and PECD data presented further on are filtered on the parent and dissociative ionization fragments centred respectively around m/z 399 and 300.

### c. Photoelectron spectroscopy and photoelectron circular dichroism: computational details

The evaluation of dynamical observables in photoionization requires the calculation of continuum wavefunctions and the dipole transition moments from the ground state. For closed-shell initial states, a generally good description is obtained with the static DFT approach, which reduces to the calculation of continuum orbitals in the static potential generated by the ground-state density, and can be further improved employing the TDDFT formalism. Note however that all TDDFT continuum calculations presented here employed an idealized $D_3$ symmetry to reduce calculation costs. As the structural deformation is very small this is believed to induce minimal changes on the photoelectron observables. The correspondence between $D_3$ and $C_2$ symmetries is $a_1 \rightarrow a$, $a_2 \rightarrow b$, $e \rightarrow a + b$.

The problematic issue is the treatment of the open-shell initial state. Two options are available. The first is to disregard it, and treat ionization at the single orbital level, by ignoring the open-shell character and simply placing three electrons, half of them as of α-spin (up) and half as of β-spin (down), into the degenerate HOMO set. The other possibility is to treat the full coupling *ab initio*, obtaining Dyson orbitals from configuration interaction (CI) wavefunctions relative to the initial and final states, and employing DFT continuum for the evaluation of the transition dipoles. We tried the latter approach utilizing small complete active space self-consistent field (CASSCF) expansions, but the results for the structure of the photoelectron spectrum were disappointing. It showed rather large couplings giving rise to wide overlapping multiplets, without an obvious correspondence with the actual spectrum.

We thus adopted the pure DFT/TDDFT approach, which is known to work reasonably well for organo-metallic complexes, as confirmed also by the good reproduction of the photoelectron spectrum (see below). To reduce the computational demands, we employed a restricted formalism, as it should not have a great impact on the transition dipoles. Both bound and continuum states and transition dipoles were computed employing a multicentric linear combination of atomic orbital (LCAO) B-spline basis set that can allow for an accurate description also of the continuum. The approach has been used several times, also for the calculation of PECD, with rather good results, and we refer to the literature for a detailed description of the method.[54-56] The basis spanned a range $R_{max}$ = 25 bohr, enough to reach the asymptotic limit, and the maximum angular momentum number was set to $L_{max}$ = 25, which ensures convergent results for the photoionization observables. The initial ground-state density which defines the DFT potential was obtained by a conventional non-relativistic LCAO calculation with the ADF program,[57] employing the full-electron STO ZORA basis from its database,[58] DZP for C, O and H, TZP for Ru, and the LB94 exchange-correlation potential,[59] which works well for photoionization calculations. All these calculations were performed at the $D_3$-symmetry Ru(acac)$_3$ structure optimized using the VWN density functional[60] with the TZP/DZP basis sets.

The computed ground-state electron density and spin density distributions may be affected by the electron delocalization error (DE) of the underlying DFT calculation.[61] Therefore, to test whether the initial LB94 density of Ru(acac)$_3$ used in the calculation of PECD is accurate, the electronic structure of the complex was also examined using a non-empirically tuned long-range corrected (LC) hybrid based on the Perdew-Burke-Ernzerhof (PBE) exchange-correlation functional,[62-64] LC-PBE0*. The tuned parametrization effectively minimizes DE[61,65] and, furthermore, it is known to reproduce ionization characteristics well from the orbital energy spectrum.[66-68] The LC-PBE0* functional affords 25% of exact exchange in the short-range limit of interelectronic distances and uses an error function range-separation parameter $\gamma^*$ of 0.1 (bohr)$^{-1}$. $\gamma^*$ was determined non-empirically based on the vertical ionization energy (IE$^{vert}$) condition, HOMO energy = -IE$^{vert}$,[66] see Table S1 in the ESI. The tuning procedure was performed with the Gaussian 09 program, revision D.01 (G09),[69] for a geometry of $\Lambda$-Ru(acac)$_3$ that was optimized using the Turbomole program, version 6.5 (TM6.5),[70-72] imposing $C_2$ symmetry, and employing the global hybrid B3LYP[73-75] density functional as implemented in TM6.5. In all these calculations, the TZVP basis set[76] was used for all atoms along with the corresponding 28-electron scalar relativistic effective core potential (ECP) for Ru.[77] B3LYP//TZVP-ECP geometry optimization was also carried out for the corresponding $D_3$-symmetry cationic system, $\Lambda$-[Ru(acac)$_3$]$^+$, with a spin triplet electronic configuration, to determine the adiabatic ionization energy (IE$^{adiab}$) of Ru(acac)$_3$ computationally.

Corresponding calculated HOMO ionization energies and comparison to the experimental determination can be found in Table 1. As can be seen, and as expected, there are large differences between the different approaches with the tuned parametrization data obtained from orbital energy clearly outperforming the other ones. The comparison showed however the consistency of LC-PBE0* and LB94 orbital composition (see discussion in Subsection d of Section *Results and discussions*), which validates the choice of the latter method for PECD calculations.

**Table 1.** Comparison of experimental and calculated IE values for the band X (see Subsection b in Section *Results and discussions*, Fig. 3 and Table 2, for the experimental determination) in Ru(acac)$_3$, in eV.

| Method | IE$^{Koop\ a}$ | IE$^{vert}$ | IE$^{adiab}$ |
|---|---|---|---|
| LB94 $^b$ | 9.83 | – | – |
| B3LYP $^c$ | 5.14 | 6.46 | 6.26 |
| LC-PBE0* $^c$ | 6.58 | 6.62 | 6.30 |
| Expt. | | 7.1 | 6.4 |

$^a$ Based on the Koopmans' theorem.
$^b$ Value obtained at VWN-optimized geometry.
$^c$ Values obtained at B3LYP-optimized geometries.

## Results and discussions

### a. Electronic circular dichroism of Ru(acac)$_3$

Fig. 2 displays ultra-violet/visible (UV/vis) absorption and ECD spectra of pure $\Delta$ and $\Lambda$ enantiomers of Ru(acac)$_3$ measured in CH$_2$Cl$_2$; see panels a and b, respectively. The UV/vis spectra demonstrate three main bands around 270, 350 and 510 nm ($\varepsilon$ between 1500 and 16500 M$^{-1}$ cm$^{-1}$), while the ECD spectrum of $\Lambda$-Ru(acac)$_3$ shows a sequence of negative-positive-negative-negative signals at 276, 347, 418 and 480 nm, with moderate intensities ($|\Delta\varepsilon|$ between 4 and 32 M$^{-1}$ cm$^{-1}$) together with a positive band of very low intensity at 545 nm ($\Delta\varepsilon$ = +1 M$^{-1}$ cm$^{-1}$). The $\Delta$-Ru(acac)$_3$ displays a mirror-image ECD signature. Note that our ECD spectra in dichloromethane exhibited similar shapes as the already reported ones, but the maximum intensity values appeared lower than those published in the supplementary information part of ref. [3] and measured in methanol. Nevertheless, 100% enantiopurity was concluded for both enantiomers of our samples since the ECD spectra always showed similar maximum values when comparing different samples from different experiments. Interested readers are referred to the ESI (Section A, Section B: Figs. S3-S10, Section C: Tables S2-S3) for the calculated UV/vis absorption and ECD spectra and their MO-to-MO contributions analyses.

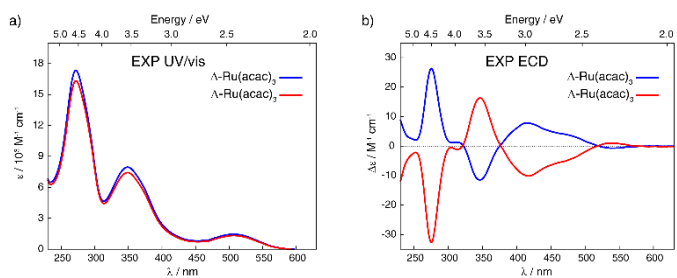

**Fig. 2.** Experimental UV/vis absorption (panel a) and ECD (panel b) spectra of $\Delta$ and $\Lambda$ enantiomers of Ru(acac)$_3$ recorded in dichloromethane (at concentrations of 5×10$^{-4}$ M).

### b. Photoelectron spectroscopy of Ru(acac)$_3$

The photoelectron spectrum of $\Delta$-Ru(acac)$_3$, obtained at a photon energy of 15 eV, is shown in Fig. 3, along with that filtered on the parent molecule (centred at m/z 399) and the fragment (m/z 300). Several features are apparent in these spectra, as evidenced in the multi-peak fit presented alongside. Because the electron energy resolution in our Velocity Map Imaging spectrometer depends strongly on the ratio between electron energy and extraction field, and also because of intrinsic spectral congestion, we have chosen to regroup the visible features of the PES into just seven bands that can be experimentally resolved over the whole photon energy range of this work. The vertical (maximum) values of these bands shown in Fig. 3 are compiled in Table 2. As will be seen in the next subsection, the PECD curves show additional structures highlighting an enhanced sensitivity of the $b_1$ parameter to the initial orbital. Therefore, the vertical positions listed in Table 2 are just indications of band maxima and do not represent ionic states. Comparison with calculations in the Subsection d will enlighten us further as to the nature and localization of the ion's electronic states, and how they compare with the experimental bands in Fig. 3.

The fact that the first IE$^{vert}$, i.e. the band maximum around 7.1 eV, is very different from the IE$^{adiab}$, which can be estimated at about 6.4 eV from the band's onset, see Table 1, hints at a rather large change of geometry upon ionization. Note that we cannot rule out the presence of a second ionic state which might be hidden under this X band, or the presence of hot bands due to the inefficient cooling of all vibrational modes, or the contribution of several conformers, so that in fact the adiabatic and vertical IE might be less separated.

As discussed previously in Subsection b of Section *Materials and methods*, a fragment at m/z 300 corresponding to the loss of an acac ligand appears at a photon energy around 12 eV. Fig. 3 also shows the PES filtered on this fragment. Its appearance energy (AE) is estimated to be 12.24 eV by a simple linear extrapolation of the onset.

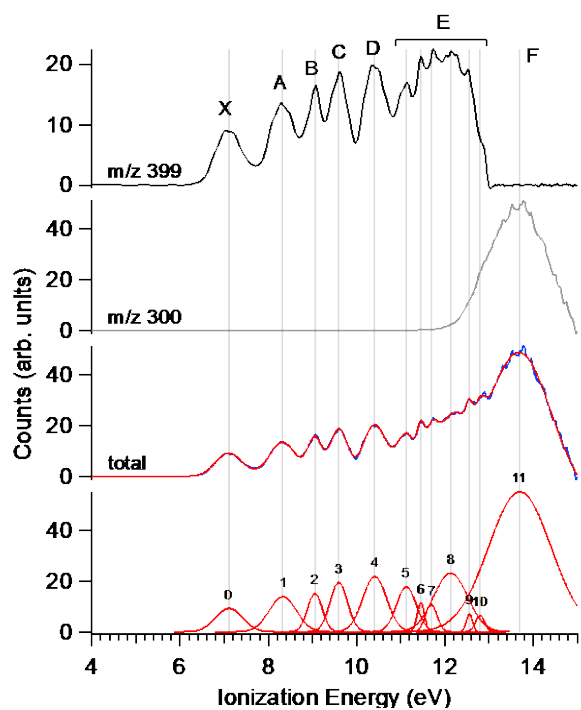

**Fig. 3.** PES of $\Delta$-Ru(acac)$_3$ recorded at a photon energy of 15 eV. The curves filtered for the parent (m/z 399, black) and fragment ions (m/z 300, grey) are given in the upper panels. The sum of the parent and fragment signals is given in the mid lower panel, together with a multi-peak Gaussian fit showing a number of visible spectroscopic features. The electronic band labels retained for the discussion are shown in the top panel.

### c. Experimental PECD of $\Lambda$- and $\Delta$-Ru(acac)$_3$ enantiomers

The PECD was recorded at several photon energies between 7.1 eV and 18 eV. Fig. 4 shows two examples of such PECD curves superimposed on the PES data, at 12 eV and 14 eV. A non-exhaustive set of experimental data, recorded at other photon energies, are also compiled in Fig. S14 of the ESI. The data display the fundamental characteristics of PECD, namely a very high intensity of the asymmetries (up to ~14% for $2b_1$, see Fig. S14c at ~9 eV ionization energy) and a clear dependence on the nature of the ionized orbital and on the electron kinetic energy (dynamical effect). As shown in Figs. 4 and S14, at the control photon energies, for which data for both enantiomers were recorded, a very good mirroring is observed within the error bars.

**Table 2.** Left: Observed bands and corresponding peak positions in the PES of Ru(acac)$_3$ (displayed in Fig. 3 and 4) and those of Ru(hfa)$_3$ from ref [80]. A shift of about 2 eV at higher energy in going from acac to hfa ligands is observed from the comparison between the two complexes. Right: DFT (LB94//B-spline) eigenvalues, corresponding molecular orbital (MO) attribution and tentative orbital assignment for Ru(acac)$_3$ (not in direct connection to the observed bands). For MOs visualization, see Fig. S11 in the ESI. All values in eV.

| band | Ru(acac)$_3$ peak position / eV | Ru(hfa)$_3$ peak position / eV | DFT | MO symmetry | orbital assignment/ nature/localization |
|---|---|---|---|---|---|
| X | 7.10 | 8.85<br>9.07 | 9.91<br>10.09 | 5e<br>2a$_1$ | metal 4d orbitals |
| A | 8.32 | 10.30 | 11.04<br>11.25 | 3a$_2$<br>4e | ligand π$_3$ orbitals |
| B | 9.05 | 11.06 | 12.11 | 3e | ligand n$_-$ orbitals, split with band C |
| C | 9.59 | 11.65 | 11.62 | 2a$_2$ | ligand n$_-$ orbitals, split with band B |
| D | 10.41 | 12.50 | 12.70<br>12.91 | 1a$_1$<br>2e | ligand n$_+$ orbitals |
| E | 11.81 { 11.13<br>11.45<br>11.69<br>12.12<br>12.61<br>12.64 | | 13.85<br><br>13.82<br>13.95<br>13.97 | 1a$_2$<br><br>1e | ligand orbitals |
| F | 13.70 | | | | |

Looking at Fig. 4 (and also Fig. S14 of the ESI), the PECD curves exhibit extra features compared to the PES data. This situation already observed in other compounds such as camphor[31] is related to the oscillatory nature of PECD curves which may switch sign when moving from a given initial orbital to a closely lying other one, leading to a better apparent resolution as compared to PES. A clear example is observed on the data recorded at a photon energy of 12 eV (Fig. 4a), which exhibit a strong feature around 7.49 eV for both enantiomers. This coincides with a small shoulder located at the same energy in the PES indicative of a different electronic state as compared to the one centered around 7.1 eV. The shoulder is only apparent in the PES at low photon energies below 13 eV (slow photoelectrons), where the Velocity Map Imaging spectrometer's energy resolution is superior (see Fig. S14 in the ESI).

In order to compare with dedicated PECD calculations, we have carried out a careful quantitative evaluation of the PECD magnitude as a function of the photon energy, or equivalently the electron kinetic energy, for the first five experimental bands (from X to D) identified in the previous section. For each band, we extract the PES-weighted average $b_1$ parameter:

$$\widehat{b_1} = \sum_{E_0-hwhm}^{E_0+hwhm} b_1 b_0 / \sum_{E_0-hwhm}^{E_0+hwhm} b_0,$$

with $b_0$ being the band PES signal, and $E_0$ and $hwhm$ corresponding to respectively the center and half-width-at-half-maximum of the considered band. The resulting $\widehat{b_1}$ versus electron kinetic energy plots are displayed in Fig. 5 for the $\Delta$ enantiomer (negated $\Lambda$ enantiomer data are also used to generate these plots). Relatively large asymmetry values are obtained for nearly all the bands, with a maximum amplitude of $\widehat{2b_1} = 0.14$ obtained for the band B, at electron kinetic energies close to 4 eV (see also Fig. S14c in the ESI). As expected, the dynamical behavior of PECD changes/oscillates dramatically with the orbital being ionized and with the kinetic energy, showing the full quantum origin of PECD. In passing, the relative enantiomeric purity was extracted from the weighted average of the dichroism over the whole PES band of each enantiomer, when available. From the 1.0 ± 0.1 average ratio found over all photon energies, both enantiomers are found to have the same enantiopurity within our error bar, in very good coherence with the ECD-derived relative enantiopurity in Subsection a above.

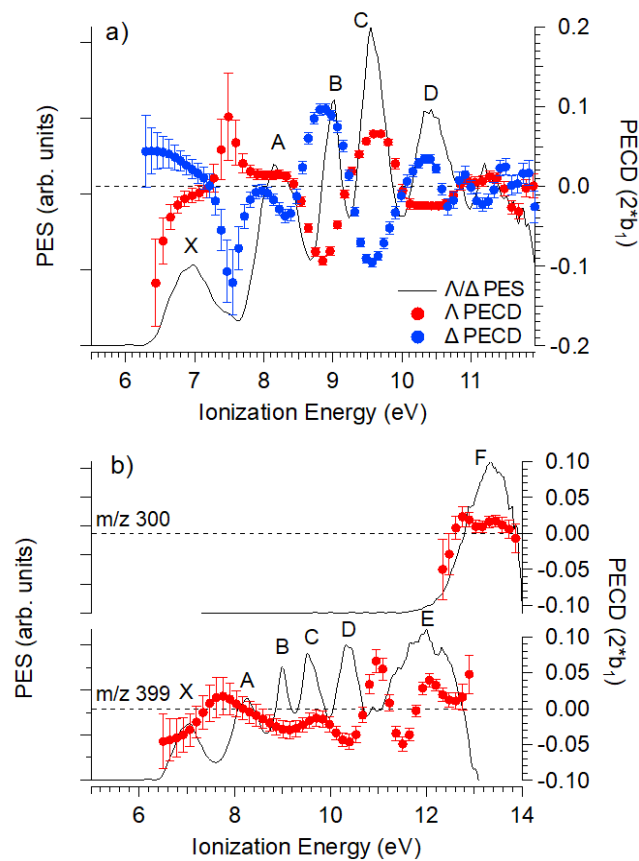

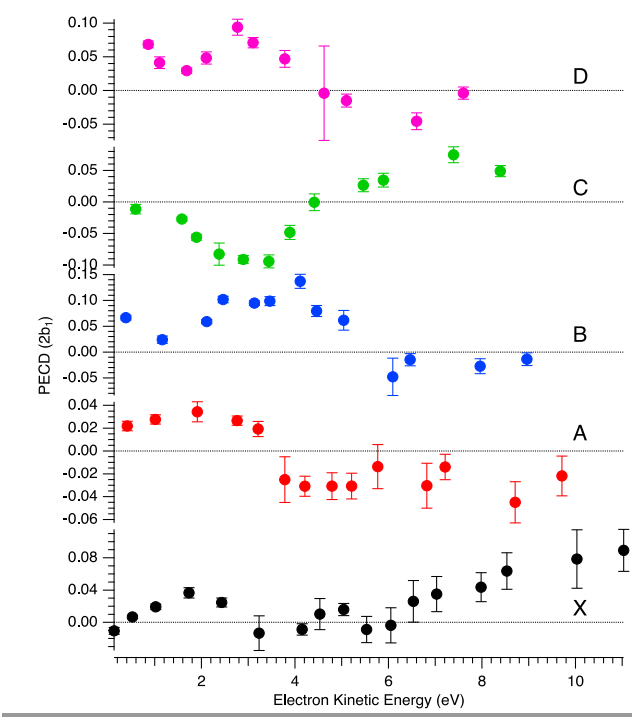

**Fig. 4.** PES (black) and PECD curves for the Δ (blue) and Λ (red) enantiomers of Ru(acac)$_3$ recorded at a) 12 eV and b) 14 eV photon energies. The electronic band X, A, B, C, D, E and F retained for the discussion already shown in Fig. 3 and compiled in Table 2 are labeled. At 14 eV, the curves for the parent (m/z 399) and fragment ions (m/z 300) are both given. At 12 eV, the only visible species is the parent Ru(acac)$_3$ cation. The statistical error bars shown are obtained by error propagation after considering an initial Poisson distribution on the photoelectron image pixel counts.

### d. Theoretical modelling, comparison with experimental PES and PECD, and discussion

Ru(acac)$_3$ is an open-shell system. If considered as the combination of an Ru$^{3+}$ ion and three acac$^-$ ligands, it corresponds to a 4d$^5$ metal configuration. This gives rise to multiplets at each orbital ionization, making a detailed analysis of the photoelectron spectrum very difficult (see below). The spectrum of acetylacetone shows two outermost ionizations at 9.0 and 9.6 eV, due to the π and n. non-bonding orbitals, followed by a composite band starting at about 12 eV.[78,79] The combination of three acac ligands around the metal gives rise formally to 4 ionizations, corresponding to two orbital couples (a$_2$ + e) in $D_3$ symmetry, but the splitting within each couple is expected to be small. The probable ground-state configuration of Ru(acac)$_3$ is a$_1^2$e$^3$ in $D_3$ symmetry (the metal 4d electrons, in a distorted $O_h$ → $D_3$ arrangement, most probably occupying t$_{2g}$ → (a$_1$ + e) orbitals) which gives a $^2E$ ground state. An *ab initio* approach gave an indication of a rather large multiplet splitting of the outermost ionization involving the 4d metal orbitals (see Subsection c of Section *Material and methods*), but gave less overall satisfactory description of the photoelectron spectrum, which we shall discuss in the following on the basis of restricted DFT results.

**Fig. 5.** Mean PECD ($\widehat{2b_1}$, see text) versus electron kinetic energy given for the Δ-Ru(acac)$_3$ enantiomer, for the first five experimental bands described in the text (see Figs. 3 and 4, and Table 2). Note that the Λ enantiomer data have been negated and are also included in these plots. The error bars reflect statistical errors (see Figure 4's caption) except those corresponding to photon energies where both enantiomers were studied (8, 11, 12 and 15 eV), which also contain systematic errors accounting for the observed deviations from a perfect mirror behaviour.

The ground state obtained from the closed-shell spin-averaged DFT LB94 calculation is ...2e$^4$ 1a$_1^2$ 3e$^4$ 2a$_2^2$ 4e$^4$ 3a$_2^2$ 2a$_1^2$ 5e$^3$, corresponding to an outer shell (n$_+$)$^6$ (n$_-$)$^6$ (π$_3$)$^6$ 4d$^5$ (a$_1$+e); the resulting DFT eigenvalues and MOs are listed in Table 2, for MOs visualization, see Fig. S11 in the ESI. The lowest ionizations are then dominated by metal 4d orbitals, while the following, π$_3$, n$_-$, n$_+$ are the acac ligand HOMO, HOMO-1 and HOMO-2 orbitals respectively. More precisely, Ru 4d orbitals participate heavily in the 5e (Ru(acac)$_3$ HOMO) and 2a$_1$ orbitals, although some contribution is spread in the lower 4e and 3e orbitals. The orbitals composing the following couple, (3a$_2$, 4e), are very close in energy and are basically a combination of ligand π$_3$ molecular orbitals, while the next couple (2a$_2$, 3e), with a larger splitting, is dominated by oxygen atomic orbitals, and correlates well with the ligands n$_-$ MOs. Then, a rather large gap separates the following group of (1a$_1$, 2e) orbitals, like in free acetylacetone, and is thus associated to the onset of the dense unresolved band of ligand valence levels, the first one being presumably a combination of ligand $n_+$ MOs.

For comparison, Figs. S12 and S13 in the ESI display selected frontier MOs resulting from the spin-unrestricted DFT calculations with the LC-PBE0* and B3LYP functionals at the $C_2$-symmetry ground-state structure of Ru(acac)$_3$. As mentioned already, LC-PBE0* is expected to provide a description of the Ru cation bonding (which is impacted by the DE) and ionization characteristics of the complex that are the best possible with hybrid Kohn-Sham DFT based on commonly used approximate functionals. Indeed, the LC-PBE0* orbital energies are well aligned with the measured ionization energy values, as illustrated for the vertical IE in Table 1 and in Table S4 of the ESI, unlike B3LYP and LB94 which produce considerable deviations. Importantly, the energetic ordering and composition of the MOs are however the same for the three functionals, with the highest-energy occupied MOs representing metal-ligand out-of-phase combinations of mainly Ru $4d_\pi$ (formally $t_{2g}$ metal orbitals of local $\pi$ symmetry with respect to the Ru–O bonds) and acac $\pi$ orbitals. In particular, we note that the degree of out-of-phase (anti-bonding) metal-ligand interactions in these MOs (reflecting the presence of in-phase, bonding overlap in corresponding occupied MOs at lower energy), is not visibly different between the spin-unrestricted B3LYP and LC-PBE0* calculations and the spin-restricted LB94 calculation. Therefore, the degree of metal-ligand covalent bonding is described similarly well with the three functionals, which means that the LB94-based assignment of the experimental bands of the PES and PECD (see below) is reliable. We note in passing that LC-PBE0* does not outperform B3LYP in reproducing the experimental UV/vis absorption and ECD spectra for the complex, despite providing better ionization characteristics (see Figs. S3 and S4 in the ESI).

Table 2 compares the DFT LB94//B-spline eigenvalues to the observed bands and corresponding positions. After scaling down the DFT eigenvalues by a very typical ~2.5-3 eV, a fairly good theory-experiment match is observed. This picture is also in close agreement with a careful and detailed analysis of the photoelectron spectrum of the analog Ru(hfa)$_3$ compound and several other metal-trischelates of acetylacetone (acac), trifluoroacetylacetone (tfa) and hexafluoroacetylacetone (hfa).[80] In this work, a comparison between acac and hfa complexes consistently showed a shift of about 2 eV towards higher energies when going from the acac- to hfa-based system. Taking this into account, we observe a good correspondence in Table 2 between the PES of Ru(hfa)$_3$ measured in ref. [80] and Ru(acac)$_3$ PES presented here (Ru(acac)$_3$ was not studied in [80]). This is also expected to correctly describe the structure of the ligand levels because, as confirmed with the *ab initio* calculations, the multiplet splitting of ligand level ionizations is almost negligible, so that metal multiplets corresponding to the same ligand orbital tend to cluster under well separated bands.

It is however much more difficult to assess the position of the metal 4d bands. As shown in Table 2, DFT calculations result in a very small splitting of the 4d = ($a_1$ + e) orbitals. On the contrary, CASSCF *ab initio* calculations predict a significant splitting (which may be partly overestimated by the lack of dynamical correlation, as it is well-known in atoms at the Hartree-Fock level) resulting in a significantly deeper $a_1$ component. The multiplet splitting of the $4d^5$ configuration should span at most a couple of eV, because of reduced inter-electronic repulsion in the molecule. However, while the lowest level is very likely associated with ionization of the 5e orbital, the position of the other metal d bands can be overlapped with those of the ligands.

Based on all this, we tentatively assign the experimental bands of the PES data as listed in Table 2 (using notations of ref. [80]). We associate with confidence the first band (X), which is rather wide, to metal 4d ionization and the higher energy bands, between 9 and 11.5 eV, to ligand ionizations. Note that the X band is quite wide and shows varying PECD profile between 6.5 and 7.5 eV, *i.e.* across the band (see Figs. 4a and S14b-d), instead of being flat as expected for a single state (neglecting kinetic energy effects), which could correspond to an unresolved multiplet structure. The position of other metal d bands is uncertain, especially considering the possible few eVs wide multiplet splitting. Together with the outermost ligand ionization, a second metal ionization could for instance lie under the 8.32 eV band (band A), which is again rather wide and which might have in fact a double structure (around 8.0 and 8.5 eV) as indicated by the corresponding modulation of the PECD curve visible in Figs. 4a and S14b. More strikingly, the very sharp feature visible on the PECD curve around 7.5 eV binding energy (see Fig. 4a and S14b-d), *i.e.* between band X and A, is a very serious hint at a specific electronic state that is very likely, considering our DFT PES analysis (see Table 2), linked to the Ru $4d^5$ multiplet structure. Given the narrow linewidth of the 9.05 eV band (band B), it seems less probable that it comprises a further metal ionization. The proposed structure is thus the following: the HOMO and HOMO-1 correspond to 4d (e + $a_1$) metal orbitals and are almost degenerate (experimental band X); they are followed by degenerate $a_2$ and e bands corresponding to ligand $\pi_3$ orbitals (experimental band A); followed by the e and $a_2$ bands which are split (experimental bands B and C) corresponding to ligand $n_-$ orbitals (oxygen $\pi$ in-plane, anti-bonding combinations); then degenerate e and $a_1$ bands corresponding to ligand $n_+$ orbitals (experimental band D, oxygen $\pi$ in-plane, bonding combinations); and finally higher, mostly unresolved, ligand ionizations. But as mentioned, it is likely that some metal multiplet bands lie at higher energies under the ligand bands (but their existence and position remain uncertain) and that several orbitals/cation states contribute to a single band.

We can see hints of the presence of such composite bands in the shape of the experimental peaks, and in the fact that PECD curves exhibit extra features compared to PES data. This is also supported by the open-shell DFT LC-PBE0*//TZVP results (see Table S4 in the ESI for the analysis of the computed MOs with reference to the experimental PES), for which the unrestricted treatment and symmetry breaking produce a large number of ionic final states (that should correspond better to the actual spectrum but simultaneously complicating the description and making a precise assignment difficult).

Importantly, the general nature of ionization envisioned from both closed-shell and open-shell approaches remain rather similar. Metal 4d ionizations dominate the lowest-energy range (corresponding to LB94 5e MOs) and participate, in combination with ligands π-orbitals, in higher-energy states (corresponding to LB94 4e, 3e orbitals). The ligand-centered MOs with large contributions from oxygen in-plane π-orbitals dominate inner ionizations (these correspond to the n orbitals in the description above). This supports the adopted approach based on the LB94 potential.

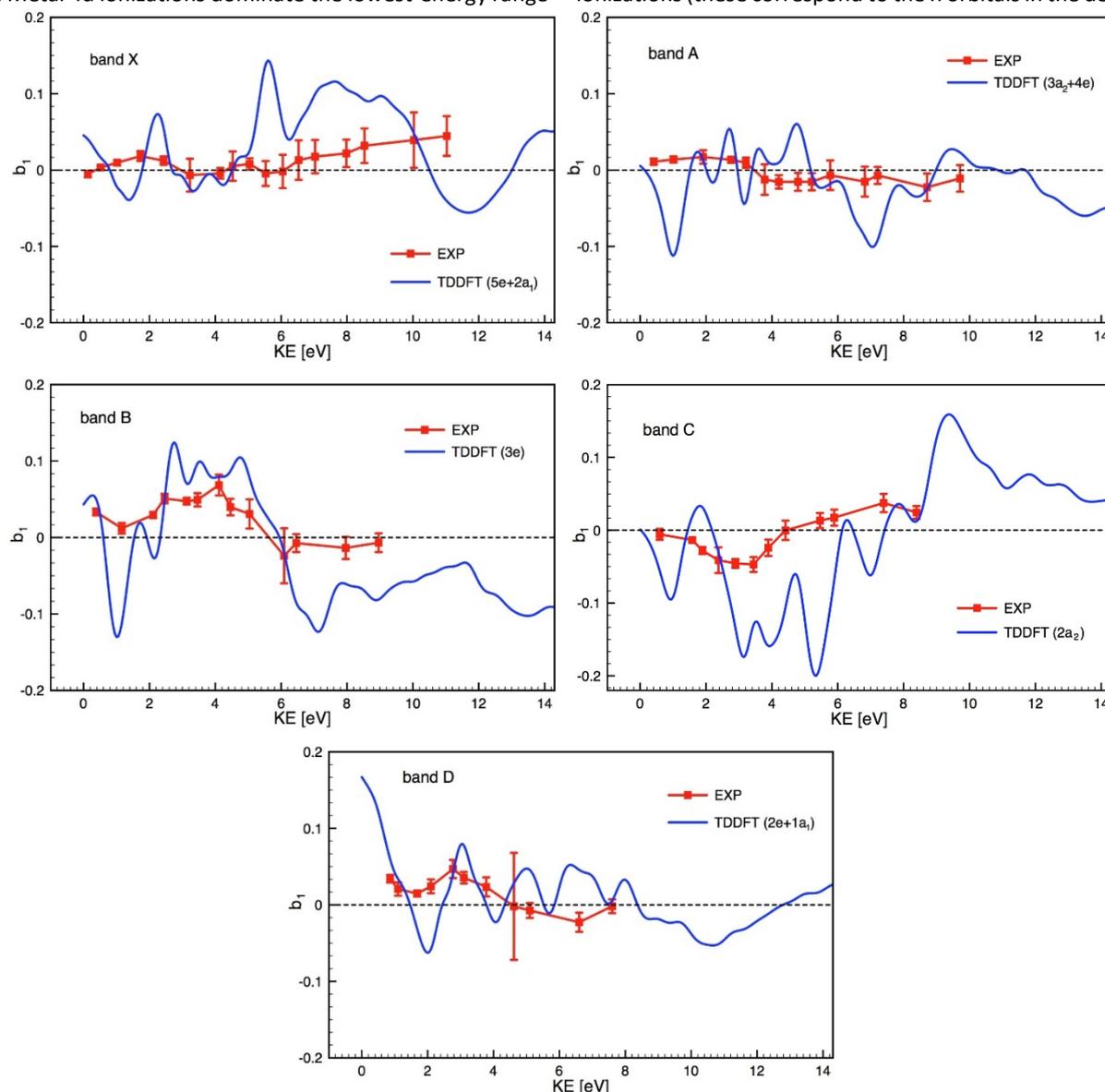

**Fig. 6.** Comparison of the experimental PECD data (EXP) and calculated TDDFT profiles of the dichroic parameter $b_1$ versus electron kinetic energy for the bands X to D identified in the PES (see Figs. 3 and 4) and listed in Table 2 (the experimental data are already shown in Fig. 5). Due to the composite nature of the experimental bands (see Table 2) each calculated profile is an average over the contributing ionization channels, weighted by the corresponding calculated cross sections.

Bearing in mind the likely mixed character of the experimental electronic bands, we compare in Fig. 6 experimental PECD data and calculated TDDFT profiles, plotted on the same scale, of the dichroic parameter $b_1$ versus electron kinetic energy for the bands X to D identified in the PES and listed in Table 2. Due to the composite nature of the experimental bands (see Table 2), Fig. 6 shows, for each band, averages over the contributing ionization channels, weighted by the corresponding calculated cross sections. Fig. S15 in the ESI also shows, for comparison, the results of DFT calculations together with experimental PECD. Results from TDDFT calculations are much more structured than DFT ones because TDDFT includes many resonances, which are generally not seen in the experiment as they are smoothed by nuclear motion, which we neglect here and simulate by artificially smoothing the TDDFT results. The agreement between experiment and theory is reasonably good, in particular for bands B, C and D, but the difficulty to unequivocally assign the photoelectron bands makes the comparison particularly challenging. As already underlined, our calculations ignore the multi-electron effects due to the multiplet structure which may well explain the poorer agreement for the two first bands, since the individual multiplet terms might have very different PECD profiles. If band A was of ligand nature only, one would expect a better agreement between theory and experiment, similar to that observed in subsequent bands. It is hard to believe that metal 4d orbitals give rise to a single band (band X) only, and so band A may well contain some underlying metal contribution that complicates the situation.

The theory has yet to progress to provide dependable reproduction and assignment of the spectra in such complex open-shell systems. Clearly a prerequisite for a detailed interpretation of PECD profiles is a sure assignment of the photoelectron bands. In this respect, complementary experimental results on cross sections, branching ratios and asymmetry parameters, up to higher energies, and including resonant ionization, may help in characterizing the orbital nature and atomic parentage of the ionic states. Such measurements, coupled with the corresponding calculations, could reveal shape resonances, which might be at the origin of the strongly varying calculated PECD feature visible around a kinetic energy of 2 eV in all photoionization channels shown in Fig. 6. Such shape resonances are known to be difficult to model in terms of PECD (see for instance ref. [33]).

Moreover, conformational flexibility of the molecule, ignored in the present study. This is known to affect the experimental/calculation interplay in a significant way. In fact, for a fair comparison, the signal of individual conformers (rotamers resulting from the presence of methyl groups) should be individually computed and then Boltzmann-averaged according to the conformer population distribution at the temperature at which the experiment is carried out.[38,44,45,81-84]

Within the composite band E, where several ionizations seem to contribute (see Fig. 3 and Table 2), comparison to theory is challenging. Nevertheless, within this band, the experimental PECD signal exhibits optima of opposite sign at binding energies of 11.1 and 11.7 eV, which we assign respectively to ionization from the $1a_2$ and $1e$ orbitals, as reported in Fig. S16 in the ESI. Although the calculated profiles appear more structured than the experimental ones, it bears some resemblance, the first being mostly positive, the second essentially negative.

In the DFT interpretation, the metal 4d orbitals (band X, and possible contribution to others) are mostly localized on the central atom, while the following orbitals are fully delocalized on the ligands, in particular combinations of oxygen 2p π and n for the bands A to D (or E). Overall, the B and C bands, corresponding to $n_-$ orbitals localized on the non-chiral ligands exhibit the most intense and oscillating curve. The X band, localized on the metal, shows a more modest PECD level but comparable to other bands, so from this data we cannot infer a direct relationship between the initial orbital localization and PECD level in the case of Ru(acac)$_3$, as already observed for other systems such as fenchone and camphor.[33] In this system however chirality is not associated with a specific stereogenic center, but is due to a global arrangement of the atoms.

## Conclusion

We carried out a complete dual experimental/theoretical study in terms of electronic structure and corresponding PECD effect on the challenging metal complex Ru(acac)$_3$. A very richly-structured PECD signature over numerous valence orbitals, for the two $\Delta/\Lambda$ enantiomers was observed in the near VUV range. A DFT-based identification of the PES features, not taking into account open-shell multiplet effects, has been proposed, suggesting an attribution of the various bands to several metal and ligand localized orbitals. Importantly, a similar PES assignment emerged also from the unrestricted calculations with the optimally tuned range-separated hybrid functional, supporting the adopted closed-shell approach based on the LB94 potential. A large number of final ionic states shown from the open-shell calculations along with the existence of very structured features in the PECD (more "resolving" than the corresponding PES profiles) point to the presence of underlying metal multiplet contributions, which the spin-restricted DFT-modelling could not predict. This probably contributes to the less than optimal agreement of the PECD experimental/TDDFT-theoretical comparison for the first two bands, while the agreement is more satisfactory for the bands attributed to non-metallic ligand-localized orbitals.

In addition, the X (and probably A) band(s), bearing most of the Ru 4d orbitals, can be spread over a broad multiplet whose individual PECD contributions may average out leading to a modest experimental PECD signal for those bands. PECD does not appear to depend on the chiral or non-chiral character of the initial orbital, nor on its localization relative to the chiral centre. This is in agreement with valence- and core-shell PECD studies in other chiral systems. In this sense, PECD is a long-range effect, created by the departing electron scattering off the whole molecular potential.

Moreover, it is worth emphasizing that the PECD signal magnitude obtained from the propeller-shaped Ru(acac)$_3$ is similar to those obtained from asymmetric carbon-based chiral species. Taking into account in addition the recent measurement of similarly intense PECD levels in axial-chirality compounds such as bi-naphthyls derivatives,[85] one can infer that PECD is a universal chiroptical effect, sensitive to any type of chirality.

This work clearly calls for more progress to be done on the theoretical side in dealing with ionization observables, and especially the very sensitive $b_1$ parameter, for open-shell heavy atom-based large systems. On the experimental side, a better electron resolution would be welcome to better separate the likely sharp metal-based multiplet spectral features from the other, overlapping orbitals.

Finally, by showing that heavy ruthenium-based chiral complexes could be handled to form a molecular beam of intact neutrals, this study opens perspectives for future high-accuracy spectroscopic measurements of chirality-related fundamental properties on such systems, like weak-force-induced parity violating energy differences between enantiomers.

## Conflicts of interest

There are no conflicts to declare.

## Acknowledgements

We gratefully acknowledge the provision of beamtime by Synchrotron Soleil under proposal numbers 20140320 and 20150103 and we thank the technical staff at Soleil for their support and for the smooth operation of the facility. A computational part of this work received financial support by the Foundation for Polish Science (Homing Plus programme co-financed by the European Regional Development Fund granted to M.S.-H.) and the National Science Foundation (grant CHE-1855470 to J.A.). J.A. acknowledges the Center for Computational Research (CCR)[86] in Buffalo, USA, and M.S.-H. thanks the PL-Grid Infrastructure and the ACC Cyfronet AGH in Krakow, Poland for providing computational resources. B.D. acknowledges the Agence Nationale de la Recherche (ANR) (grant ANR-15-CE30-0005-01). A.P. gratefully acknowledges Dr. Nicola Quadri for fruitful comments and discussions.